\documentclass[11pt,a4paper]{article}
\pdfoutput=1
\usepackage{jcappub}

\def\Teq{{{T_{\rm eq}}}}

\title{Tidal streams from axion miniclusters and direct axion searches}

\author{Peter Tinyakov$^a$}
\affiliation{$^a$Universite Libre de Bruxelles, Service de Physique Theorique, CP225, 1050, Brussels, Belgium}
\author{Igor Tkachev$^b$}
\affiliation{$^b$Institute for Nuclear Research of the Russian Academy of Sciences, Moscow 117312, Russia}
\author{Konstantin Zioutas$^c$}
\affiliation{$^c$University of Patras, Greece and CERN, Geneve, Switzerland}

\abstract{
In some axion dark matter models a dominant fraction of axions resides in
dense small-scale substructures, axion miniclusters. A fraction of these
substructures is disrupted and forms tidal streams where the axion density may
still be an order of magnitude larger than the average.  We discuss
implications of these streams for the direct axion searches. We estimate the
fraction of disrupted miniclusters and the parameters of the resulting streams,
and find that stream-crossing events would occur at a rate of about $1/(20
{\rm yr})$ for 2-3 days, during which the signal in axion detectors would be
amplified by a factor $\sim 10$. These estimates suggest that the effect of
the tidal disruption of axion miniclusters may be important for direct axion
searches and deserves a more thorough study.}

\begin{document}
\maketitle
\flushbottom

\section{Introduction}

In a wide variety of axion Dark Matter (DM) models, a sizable (or even
dominant) fraction of axions is confined in a very dense axionic clumps, or
miniclusters, with masses $M \sim 10^{-12}M_{\odot}$. The axion miniclusters
originate from specific density perturbations which are a consequence of
non-linear axion dynamics around the QCD epoch \cite{Hogan:1988mp,Kolb:1993zz,Kolb:1993hw}.
There may be $\sim 10^{24}$ of such miniclusters in the Galaxy, their density
in the Solar neighborhood being $\sim 10^{10} \rm ~pc^{-3}$. Typical
miniclusters have radius of $\sim 10^7 ~\rm km$ and the density $\sim
10^8$~GeV\,cm$^{-3}$. During a direct encounter of the laboratory with such a
minicluster the local axion density increases by a factor of $10^8$ for about
a day. That would create a very strong signal in the tuned detectors devoted
to direct axion searches. However, direct encounters with the Earth would
occur only once in $\sim 10^5 ~\rm years$~\cite{Kolb:1994fi}.

Over the lifetime of the Galaxy, the axion miniclusters may be tidally
disrupted forming tidal streams. In this paper we discuss possible
phenomenological consequences of these structures for the direct axion
searches. As we will argue, the tidal streams have a much larger volume than
the original miniclusters, which boosts the encounter rate. At the same time
they may still be dense enough to produce a sizable signal in axion
detectors.

A similar idea has already been discussed in a general context of weakly
interacting massive particles (WIMPs). In any cold dark matter model
(including WIMPs and axions) halos are formed, by a standard gravitational
instability, on all scales from galaxies down to a free-streaming scale,
leading to structure formation from primordial density perturbations. In case
of WIMPs, the smallest halos have masses around $M \sim 10^{-7} M_\odot$ as
set by the free streaming scale in a typical WIMP model
\cite{Hofmann:2001bi,Berezinsky:2003vn,Berezinsky:2014wya}.  For the axion DM
the minihalos form down to even smaller masses $M \sim 10^{-12} M_\odot$,
which is the mass of all axions inside the horizon at the epoch when the
axion oscillations commence. This process has been numerically modeled both
for WIMPs and axions in Ref.~\cite{Diemand:2005vz} in the mass range $
10^{-6}\,M_{\odot} ~\lesssim~ M ~\lesssim~ 10^{-4}\,M_{\odot}$. For minihalos
with $M \sim 10^{-6} M_\odot $ their density in the Solar neighborhood after
formation was found to be $\sim 500 \rm ~pc^{-3}$, the direct encounter with
the Earth would occur once in $10^4$ years, and during the encounter the DM
density would increase by a factor of 100 over the average for about 50
years. However, virtually all minihalos which are crossing Solar neighborhood
are tidally destroyed long before the present
epoch~\cite{Zhao:2005py,Berezinsky:2005py,Green:2006hh,Goerdt:2006hp,Berezinsky:2014wya}
producing a density field not interesting for the direct detection of
DM~\cite{Schneider:2010jr}.  The situation may be different for axion
miniclusters due to an additional, different formation mechanism that leads to
smaller and denser clumps, the axion miniclusters.

\section{Axion Miniclusters}

Unlike other forms of DM, axions may develop density fluctuations of order one
(or even larger) already at early times, prior to radiation-matter
equality. This happens due to a peculiar (non-linear) axion dynamics at the
epoch when axion oscillations commence. Let us denote by $\Phi =
\delta\rho_a/\rho_a$ the energy density contrast in the axion filed at the end
of this epoch.  Consider the class of models where Peccei-Quinn symmetry is
restored during reheating or preheating after inflation.  Then the ratio
$\theta = a/f_a$, where $a$ is the axion field and $f_a$ the axion decay
constant, can take any value from one horizon volume to another, while being 
smooth within a given horizon.  When the axion oscillations start due to
development of a non-zero axion potential
$V(\theta)=m_a^2f_a^2[1-\cos(\theta)]$ at the beginning of the QCD epoch, the
field fluctuations $\delta\theta \sim 1$ are transformed into density
fluctuations $\Phi\sim 1$.  Numerical investigations of the dynamics of the
axion field at this epoch
\cite{Kolb:1993zz,Kolb:1993hw,Kolb:1994fi,Kolb:1995bu} have shown that the
non-linear effects lead to clumps with $\Phi$ much larger than unity, possibly
as large as several hundred. These regions separate from cosmological
expansion at temperature $T \simeq \Phi\Teq$, where $\Teq$ is the temperature
of equal matter and radiation energy densities.  This results in a final
minicluster density today given by~\cite{Kolb:1994fi}
\begin{equation}
\label{rhofl}
\rho_{\rm mc} \simeq 7\times10^6 \, \Phi^3 (1+\Phi)\, \rm GeV / cm^{3}.
\end{equation}
This should be compared to the mean DM density in the Solar
neighborhood in the Galaxy, $\bar{\rho} \approx 0.3 \; {\rm GeV/cm}^3$.

The scale of minicluster masses is set by the total mass $M_1$ of 
axions within the
Hubble radius at a temperature around $T \approx 1$~GeV when axion
oscillations start,
\begin{equation}
M_{1} \approx 10^{-12}M_\odot .
\label{Eq:mass}
\end{equation}
Numerical integration of Ref~\cite{Kolb:1995bu} shows that a spectrum
of minicluster masses is produced that is compact and concentrated
around a sizable fraction of $M_1$, so that $0.1 M_1\lesssim M_{\rm
  mc} \lesssim M_1$. In what follows, however, we do not rely on this
spectrum. Instead, we assume a single value of $M_{\rm mc}=
10^{-12}M_\odot$ for simplicity, except for our final result,
Fig.~\ref{fig:2}, which is presented for two bracketing values $M_{\rm
  mc}= 10^{-12}M_\odot$ and $M_{\rm mc}= 10^{-13}M_\odot$.

The minicluster radius as a function of $\Phi$ and  $M_{\rm mc}$ is
\begin{equation}
R_{\rm mc} \approx  \frac{3.4 \times 10^{7}}
{\Phi \left(1+\Phi\right)^{1/3} } 
\left(\frac{M_{\rm mc} }{10^{-12} M_\odot}\right) ^{1/3} 
{\rm km} \, ,
\label{R}
\end{equation}
while the typical virial velocity in the minicluster, $v_{\rm mc}^2 = GM_{\rm
  mc}/R_{\rm mc}$ is given by
\begin{equation}
v_{\rm mc} \approx 2.1\times 10^{-10} \; \Phi ^{1/2}\left(1+\Phi\right)^{1/6}  
\left(\frac{M_{\rm mc} }{10^{-12} M_\odot}\right)^{1/3} .
\label{v}
\end{equation}

\begin{figure}
\includegraphics[width=1\columnwidth]{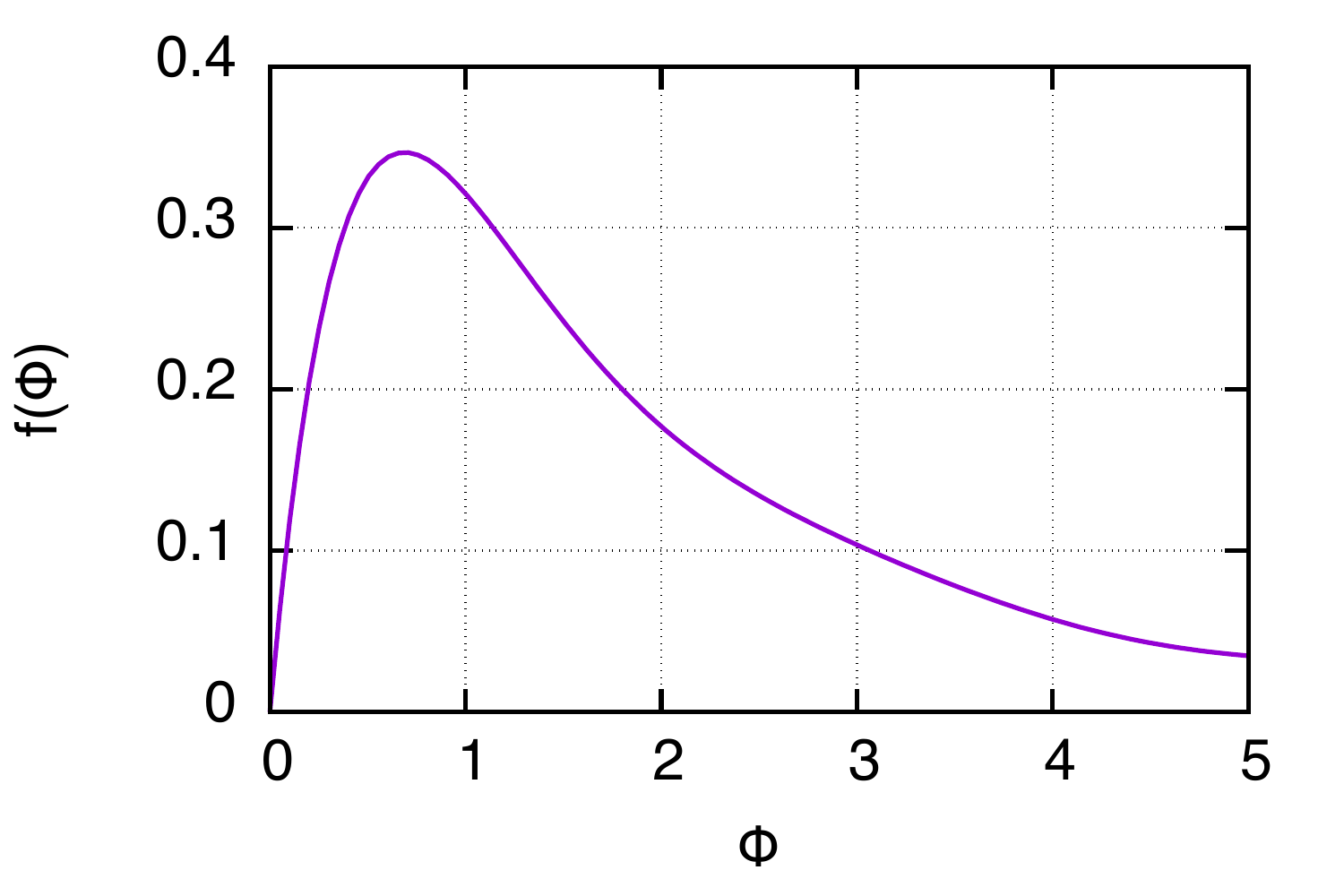}
\caption{\label{fig:1} 
Mass fraction $f(\Phi)$ of axions in miniclusters with a given value of $\Phi$.  }
\end{figure}

The distribution of miniclusters in values of $\Phi$ has been calculated
numerically in Ref.~\cite{Kolb:1995bu} neglecting contribution from the decaying 
axion strings and domain walls.  This distribution is summarized in
Fig.~\ref{fig:1} which shows a differential probability $f(\Phi)$ to find an
axion in a clump with a given value of $\Phi$, as a function of
$\Phi$. According to this plot, 70\% of all axions are in miniclusters with
$\Phi > 1$. 
Contribution from the decay of the 
domain wall -string network is likely to modify this distribution only quantitatively. 
Indeed, since i) the axion string network decays completely at the epoch when 
the axion oscillations start ii)  the process is
highly inhomogeneous being dominated by the horizon scale iii) produced axions 
are non-relativistic, see e.g. \cite{Wantz:2009it},  one can expect that the change in the 
miniclusrer distribution will not be qualitative.

\section{Tidal destruction of miniclusters by stars}

Small-scale clumps in the Galaxy may be tidally disrupted by the
gravitational field of the halo, of the disc and in encounters with
stars. These processes were studied by many authors, for a review  see
\cite{Berezinsky:2014wya}. The axion miniclusters are too dense to be disrupted
by the halo gravitational field. Here we estimate the rate at which
they are disrupted in stellar encounters, which is the dominant
process for the relevant range of masses.

When a minicluster traverses the stellar field, interactions with
individual stars increase the velocity dispersion of dark matter particles
and reduce the clump's binding energy.  Following
Refs.~\cite{Goerdt:2006hp,Schneider:2010jr} we introduce a critical
impact parameter, $b_c$, such that for $b < b_c$ a single encounter is
sufficiently strong to unbind the minicluster.  This parameter was
found to be \cite{Goerdt:2006hp,Schneider:2010jr}
\begin{equation}
b_{c}^2 \approx \frac{GM_{s}R_{\rm mc}}{v_{\rm rel}\, v_{\rm mc}} , 
\label{eq:b_c}
\end{equation}
where $v_{\rm rel}$ is the relative velocity of a minicluster and a star, and 
$M_s$ is the star mass. Therefore, the disruption cross section in such a
``strong'' encounter is $\pi b_c^2$ and is proportional to the star
mass. The break-up probability in a single passage of a minicluster
through the Galactic disc is
\begin{equation}
p_s=\pi {GR_{\rm mc}\over v_{\rm rel}\,  v_{\rm mc} } S,
\label{eq:p_s}
\end{equation}
where $S$ is the column mass density of stars. Note that, because of
eq.~(\ref{eq:b_c}), only the total star mass density enters this
expression, while the distribution of stars in masses drops out. Also,
$p_s$ does not depend on minicluster mass, cf. eqs. (\ref{R}) and
(\ref{v}).

According to Ref.~\cite{Kuijken:1989hu} the stellar contribution to
the column density is $S_\bot \approx 35 \; \rm M_\odot pc^{-2}$ in
the direction orthogonal to the disc, and correspondingly larger for
the inclined directions.  We are interested in minicluster
trajectories which pass through the vicinity of the Earth. Assuming
for the sake of the estimate that such trajectories are distributed
isotropically, the integral over the angles diverges logarithmically
when approaching the direction parallel to the disk; it has to be cut
at the typical length of the trajectory that lies completely in the
disk, for which we take $O(10{\rm kpc})$.  Integration over directions then gives 
\[
 S \approx 4S_\bot. 
\]
The cumulative effect of multiple non-disruptive encounters with $b >
b_c$ increases the disruption probability by a factor of about
two~\cite{Goerdt:2006hp}, $P = 2p_s$.

To estimate the total probability of a minicluster disruption until
the present epoch we multiply the disruption probability in a single
crossing of the disk by the number of crossings over the whole
history of the Galaxy, $n\sim 100$. Assuming $v_{\rm rel} = 10^{-3}$ in eqs.~(\ref{eq:p_s}) we find
\begin{equation}
P(\Phi) = 0.022 \left( {n\over 100} \right) \Phi ^{-3/2}\left(1+\Phi\right)^{-1/2} . 
\label{eq:sp}
\end{equation}
Only a few percent  of miniclusters with $\Phi \sim
1$ are destroyed, the rest remain intact. However, even this small fraction of
disrupted miniclusters may be important for direct axion searches.

\section{Axion streams and implications for direct axion searches.}

Debris from disrupted miniclusters form tidal streams along the minicluster
trajectory. 
For orbits with small ellipticity the streams are essentially
one-dimensional structures of large length $L$ and cross-section
comparable to the initial clump size.  Recently this
process was modeled for clumps with $M \sim 10^{-6} M_\odot $, see
  Ref.~\cite{Angus:2006vp,Schneider:2010jr}, while for the general
theory and arbitrary orbits see
e.g. Refs. \cite{Helmi:1999ks,Lux,Bovy:2014yba}.

Following Ref.~\cite{Schneider:2010jr} we estimate the length of the tidal
streams due to the orbiting process as $L \sim v_{\rm mc} t$, where $t$ is the
age of a stream.  In this approximation, when a minicluster is disrupted and
forms a tidal stream, its volume grows by a factor $v_{\rm mc} t / R_{\rm
  mc}$. The axion density in the stream drops accordingly,
\begin{equation}
\rho_{\rm st} (\Phi)= \rho_{\rm mc} {R_{\rm mc}\over v_{\rm mc} t},
\label{eq:rho-st}
\end{equation}
and does not depend upon minicluster mass.

The time $\tau$ to cross  such a stream by an observer moving with the
relative velocity $v_{\rm rel}$ does not depend on the stream age,
\[
\tau (\Phi, M_{\rm mc}) = {2R_{\rm mc}}/v_{\rm rel} \approx
\]
\begin{equation}
\approx{62\, \rm
  hr}\; {\Phi^{-1} \left(1+\Phi\right)^{-1/3}} 
\left({M_{\rm mc}}/{10^{-12} M_\odot}\right) ^{1/3}
. 
\label{eq:tau}
\end{equation}
In the context of axion detection, this time interval corresponds to a period
of high signal in the detector. As this signal is proportional to the axion
density, it is convenient to introduce the {\em amplification factor}
\begin{equation}
A = {\rho_{\rm st}} /{\bar\rho}.
\label{eq:A}
\end{equation}
The value $A=0$ corresponds to no signal amplification during the string crossing event. 

If the disrupted miniclusters had the same axion density and were disrupted at
the same time, all resulting streams would have the same axion
density. Knowing the volume filling factor of these streams in the Galaxy
$\epsilon$ and the crossing time $\tau$, the frequency of the stream-crossing
events would then be
\begin{equation}
\nu = {\epsilon\over \tau}. 
\label{eq:nu}
\end{equation}
The meaning of $\nu$ is that the mean number of stream-crossings  in a
given time interval $\Delta t$ is $N =\nu \Delta t$. It is important to note
that $\nu$ is proportional to $\epsilon$ and therefore contributions to $\nu$
of different types of miniclusters can be simply added.

In a realistic situation the miniclusters are distributed in $\Phi$ (and,
therefore, in density) as follows from Fig.~\ref{fig:1}.  The disruption
probability of a minicluster also depends on $\Phi$,
cf. eq.~(\ref{eq:sp}). Moreover, the miniclusters of a given $\Phi$ can be
disrupted at different times, which are distributed uniformly from the moment
of the Galactic stellar disc formation until present; they would then produce
streams of a different density. The contribution into the stream-crossing rate
of streams of a given amplification factor $A$ originated from miniclusters of
given $\Phi$ can be written as follows:
\begin{equation}
d\nu = {P(\Phi) f(\Phi)\over \tau(\Phi,M_{\rm mc})} 
{a(\Phi)\over A^3} dA d\Phi.
\label{eq:d_nu}
\end{equation}
Here $f(\Phi)$ is the mass fraction in miniclusters, Fig.~\ref{fig:1},
$P(\Phi)$ and $\tau(\Phi,M_{\rm mc})$ are given by eqs.~(\ref{eq:sp}) and
(\ref{eq:tau}), and $a(\Phi)= \rho_{\rm st}(\Phi)/\bar\rho$ is the
amplification factor corresponding to the minicluster with given
$\Phi$ disrupted right after the Galactic disk formation as given by
eq.~(\ref{eq:rho-st}) with $t$ equal to the age of the disk.  
The physically relevant quantities are obtained by integrating
eq.~(\ref{eq:d_nu}) over appropriate range. 
It is convenient to introduce $\nu(A)$ as
a frequency of  stream crossings 
with the amplification larger than the given value $A$. To obtain
$\nu(A)$, eq.~(\ref{eq:d_nu}) has to be integrated over $A$ from $\max
(A, a(\Phi))$ to $\rho_{\rm mc}/ {\bar\rho}$ as 
determined by eq. ~(\ref{rhofl}), and over $\Phi$. 


\begin{figure}
\includegraphics[width=1\columnwidth]{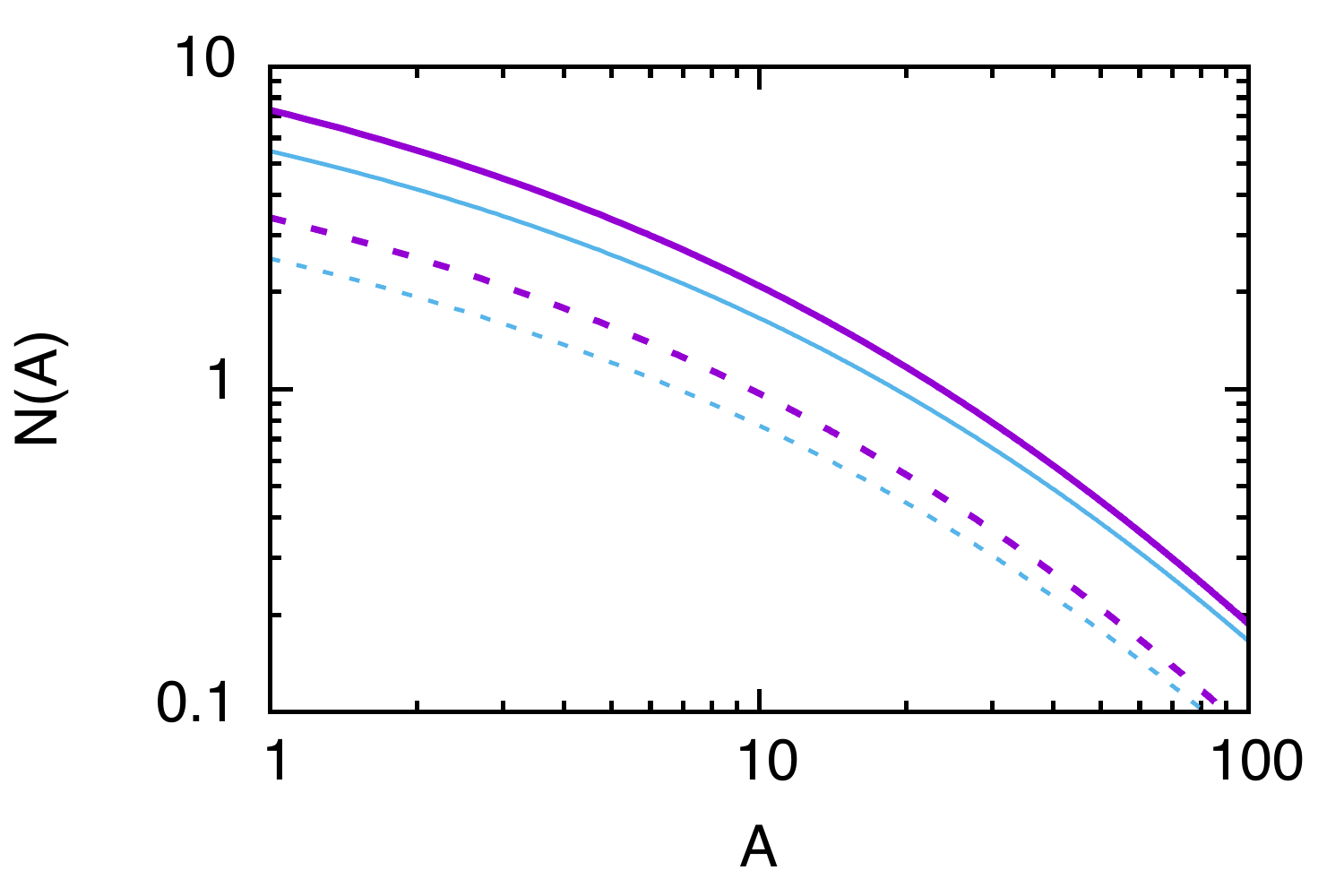}
\caption{\label{fig:2} Mean number of encounters with axion streams producing
  amplification factor larger than $A$, as a function of $A$. Twenty year
  observation interval is assumed. Solid and dashed curves correspond to
  $M_{\rm mc}$ equal to $10^{-13} M_\odot$ and $10^{-12} M_\odot$,
  while thick dark and thin light lines to 12 and 10 Gyr age of
  the Galactic disc, respectively.}
\end{figure}

To characterize the prospects of axion detection in encounters with
tidal streams, we present in Fig.~\ref{fig:2} the mean number of
stream encounters $N(A) = \nu(A) \Delta t$, where 
the observation time interval was taken to be $\Delta t = 20$~yr. 
The upper pair of curves (solid lines) corresponds to $M_{\rm
  mc}=10^{-13}M_\odot$, while the lower pair (dashed lines) to $M_{\rm
  mc}=10^{-12}M_\odot$. Within each pair the thick curve corresponds
to the age of the  stellar Galactic disc of 12~Gyr, and the thin one to 10~Gyr.  As one
can see from the plot, for the experiments sensitive to the
amplification factors $A\sim 10$ the probability of detection is close
to 1.

\section{Conclusions}

To summarize, we have considered the effect of the tidal streams that are
produced in disruptions of axion miniclusters by the encounters with the stars
in the Galactic disk. Despite only a few percent of the miniclusters get
disrupted by the present time, the much larger volume of the resulting tidal
streams as compared to the miniclusters themselves boosts the encounter rate
to an acceptable value of 1 in $\sim 20$~yr, while still providing a sizeable
amplification of the signal $A\sim 10$.

Our estimates have been made in a particular axion DM model where the 
formation of clumps and their parameters have been studied numerically \cite{Kolb:1994fi} neglecting 
the contribution of decaying axion strings and domain walls. This calculation 
should be improved by including the effects of these structures. 
We also have made a number of simplifying
assumptions: we assumed that all miniclusters have the same mass; we
estimated the parameters of the tidal streams rather crudely and
neglected their possible dependence on the parameters of the clump
orbit; the orbits themselves were assumed to have isotropic
distribution around the Earth.  A more accurate calculation is clearly
needed to take these effects into account. 

However, already the present estimate suggests that the effect of the tidal
disruption of axion miniclusters may be important for the direct relic axion
searches. In particular, our results give extra support for a wide band detection systems. 
Such systems will not replace of course, but may complement usual narrow axion mass scanning
haloscopes. Generically, the latter will not be tuned to the true axion mass during encounter 
with the region of higher axion density, and the event will be missed. In addition, if miniclusters are abundant, the expected signal between encounters is smaller as compared to the one produced by mean density of axions.  This would require extra sensitivity for a resonance  detector to exclude given mass range reliably. Also, the ADMX experiment~\cite{Asztalos:2009yp} is slowly scanning an order of magnitude mass range around $10^{-5}$ eV, while e.g. wide band dish antenna similar to suggested in Ref.~\cite{Horns:2012jf} may be operational in a much wider range simultaneously, including  higher frequencies corresponding to models where axion minicluster formation is expected\footnote{The threshold energy of the dish antenna detector depends inversely proportional on the antenna diameter. For a 10 cm dish size the threshold is about $10^{-4}$ eV \cite{Lindner}}. Wide band detectors might not have sensitivity high enough to detect mean axion background, but may produce detectable signal during encounter event. Clearly, it is not possible to  rely on a once-in-twenty-year transient event to identify dark matter. One would need to have at least two such detectors separated geographically. In any case, if axions will be ever discovered, the subsequent search for the rare events, such as tidal stream crossings, will provide valuable information both for the Galactic and the early Universe  histories.

\section{Acknowledgments}

We are grateful to V. Dokuchaev  for useful  discussions. One of us (K.Z.) wishes to thank Sergio Bertolucci for interesting discussions on dark matter.
This work was supported by the Russian Science Foundation grant 14-22-00161.

\section{Bibliography}

\begin{footnotesize}

\end{footnotesize}


\end{document}